\documentclass[a4paper,fleqn,sort&compress]{cas-dc}

\usepackage[numbers]{natbib}
\usepackage{amsmath}
\usepackage{bm}
\usepackage{colortbl}
\usepackage{xcolor}
\usepackage{color}
\usepackage{soul}
\usepackage{bbding}
\usepackage{multirow}
\usepackage{multicol}
\usepackage{booktabs}
\definecolor{light-gray}{gray}{0.90}
\definecolor{mydarkgreen}{RGB}{0,150,0}

\def\tsc#1{\csdef{#1}{\textsc{\lowercase{#1}}\xspace}}
\tsc{WGM}
\tsc{QE}
\tsc{EP}
\tsc{PMS}
\tsc{BEC}
\tsc{DE}

\begin{document}
\let\printorcid\relax
\let\WriteBookmarks\relax
\def\floatpagepagefraction{1}
\def\textpagefraction{.001}

\shorttitle{MOSformer: Momentum Encoder-based Inter-slice Fusion Transformer for Medical Image Segmentation}

\shortauthors{D.-X. Huang et~al.}

\title [mode = title]{MOSformer: Momentum Encoder-based Inter-slice Fusion Transformer for Medical Image Segmentation}

\author[1,2]{De-Xing Huang}
\ead{huangdexing2022@ia.ac.cn}
\credit{Data curation, Methodology, Writing - original draft}

\author[1,2]{Xiao-Hu Zhou}
\cormark[1]
\ead{xiaohu.zhou@ia.ac.cn}
\credit{Conceptualization, Funding acquisition, Writing - review and editing}

\author[1,2]{Mei-Jiang Gui}
\credit{Resources, Funding acquisition}

\author[1,2]{Xiao-Liang Xie}
\credit{Funding acquisition}

\author[1,2]{Shi-Qi Liu}
\credit{Resources}

\author[1,2]{Shuang-Yi Wang}
\credit{Resources}

\author[1,2]{Zhen-Qiu Feng}
\credit{Resources}

\author[3]{Zhi-Chao Lai}
\credit{Resources}

\author[1,2]{Zeng-Guang Hou}
\cormark[1]
\ead{zengguang.hou@ia.ac.cn}
\credit{Supervision}

\affiliation[1]{organization={State Key Laboratory of Multimodal Artificial Intelligence Systems, Institute of Automation, Chinese Academy of Sciences},
city={Beijing},
postcode={100190},
country={China}}

\affiliation[2]{organization={School of Artificial Intelligence, University of Chinese Academy of Sciences},
city={Beijing},
postcode={100049},
country={China}}

\affiliation[3]{organization={Department of Vascular Surgery, Peking Union Medical College Hospital},
city={Beijing},
postcode={100730},
country={China}}

\cortext[cor1]{Corresponding authors.}

\begin{abstract}
Medical image segmentation takes an important position in various clinical applications. 2.5D-based segmentation models bridge the computational efficiency of 2D-based models with the spatial perception capabilities of 3D-based models. However, existing 2.5D-based models primarily adopt a single encoder to extract features of target and neighborhood slices, failing to effectively fuse inter-slice information, resulting in suboptimal segmentation performance. In this study, a novel momentum encoder-based inter-slice fusion transformer ({\tt MOSformer}) is proposed to overcome this issue by leveraging inter-slice information from multi-scale feature maps extracted by different encoders. Specifically, dual encoders are employed to enhance feature distinguishability among different slices. One of the encoders is moving-averaged to maintain consistent slice representations. Moreover, an inter-slice fusion transformer (IF-Trans) module is developed to fuse inter-slice multi-scale features. {\tt MOSformer} is evaluated on three benchmark datasets (Synapse, ACDC, and AMOS), achieving a new state-of-the-art with 85.63\%, 92.19\%, and 85.43\% DSC, respectively. These results demonstrate {\tt MOSformer}'s competitiveness in medical image segmentation.
\end{abstract}

\begin{keywords}
Medical Image Segmentation \sep Momentum Encoder \sep Inter-slice Fusion \sep Transformer
\end{keywords}

\maketitle

\section{Introduction} \label{sec:s1}
Medical image segmentation plays a crucial role in numerous clinical applications, such as computer-aided diagnoses~\cite{zhou2020real},~\cite{antonelli2022medical}, image-guided interventions~\cite{li2021unified,li2021real,han2024respiratory,huang2025real,chen2025gvm}, and surgical robotics~\cite{zhang2024ai,li2023three,zhou2022learning,zhang2025novel}. {\tt UNet}~\cite{ronneberger2015u} and its variants~\cite{zhou2019unet++,huang2020unet,ni2022surginet,huang2024spironet} have been widely used in this field, achieving tremendous success in different medical imaging modalities. However, accurate and efficient segmentation of 3D medical images still remains a non-trivial task~\cite{tajbakhsh2020embracing}.

Current mainstream segmentation methods can be classified into two categories: 2D-based and 3D-based methods~\cite{zhang2022bridging}. 2D-based methods split 3D images into 2D slices and segment them individually, while 3D-based methods divide 3D images into smaller patches and then segment these patches individually. Despite impressive performance achieved by state-of-the-art methods~\cite{azad2024medical}, they still exhibit some limitations. Most 2D-based methods focus on architecture design to enhance intra-slice representations for better performance, such as incorporating attention modules~\cite{mou2021cs2},~\cite{roy2022lwmla} or adopting transformers~\cite{cao2022swin},~\cite{chen2024transunet}. However, these methods overlook inter-slice cues, which are also crucial for accurate segmentation. In contrast, 3D-based methods can capture intra- and inter-slice information for segmentation but demand substantial GPU memory and computational resources. Additionally, they tend to perform poorly in images with anisotropic voxel spacing since they are primarily designed for 3D images with nearly isotropic voxel spacing~\cite{cciccek20163d},~\cite{yu2017automatic}.

In order to combine the advantages of 2D-based and 3D-based methods, some studies have been done to explore 2.5D-based segmentation models~\cite{zhang2022bridging}. The main idea of these methods is to fuse inter-slice (neighborhood slices) information into 2D-based models when segmenting specific slices (target slices). The most direct way to achieve inter-slice fusion is by concatenating slices as multi-channel inputs. However, it is inefficient, making it challenging for models to extract useful features for the target slice~\cite{zhang2022bridging}. Therefore, some studies focus on exploring ``smart'' ways of inter-slice fusion. Most of them formulate 2D slices as time sequences and adopt recurrent neural network (RNN)~\cite{chen2016combine}, transformers~\cite{yan2022after},~\cite{hung2022cat} or attention mechanisms~\cite{zhang2020sau} to fuse inter-slice information.

\begin{figure}[htbp]
\centering
\centerline{\includegraphics{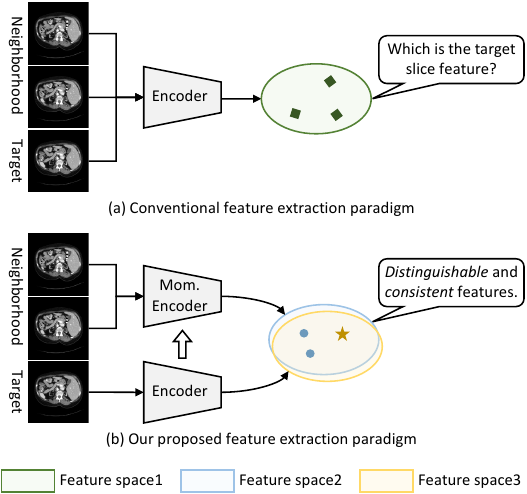}}
\caption{Comparison between conventional feature extraction paradigm of 2.5D-based segmentation models and our proposed paradigm. (a) Conventional approaches use a single encoder to extract features of all slices. Therefore, target slices and neighborhood slices share the same feature space. (b) Our proposed paradigm adopts dual encoders to extract features of target and neighborhood slices, respectively. Momentum update is used in the neighborhood slice encoder. Hence, feature spaces of target and neighborhood slices are distinguishable and consistent. (Mom.: Momentum.)}
\label{fig:1}
\end{figure}

While current 2.5D-based methods have achieved impressive segmentation results, they struggle with distinguishing individual slices during inter-slice fusion, and consequently fail to learn reliable inter-slice representations essential for accurate segmentation~\cite{zhang2022bridging}. The root of this problem lies in the use of a single encoder to process all input slices, resulting in the same feature distributions across the feature space, as shown in Fig.~\ref{fig:1} (a). For example, the features of the $i$-th slice remain identical whether it is considered as the target slice or the neighborhood slice. This indistinguishability becomes problematic in scenarios where consecutive slices, such as the $i$-th and the $(i+1)$-th, are target slices, respectively. Models fail to differentiate the $i$-th slice's features as belonging to the target or the neighborhood slice, thereby hampering the extraction of valuable inter-slice information for segmentation.

To address the above issue, a novel 2.5D-based segmentation model, {\tt MOSformer}, \underline{\tt MO}mentum encoder-based inter-\underline{\tt S}lice fusion trans\underline{\tt former} is proposed to effectively leverage inter-slice information for 3D medical image segmentation. {\tt MOSformer} follows the design of the U-shaped architecture~\cite{ronneberger2015u}. In order to enhance feature distinguishability of each slice, dual encoders are utilized in our model, with one for target slices and the other for neighborhood slices. Parameters of the target slice encoder are updated by back-propagation, and parameters of the neighborhood slice encoder are updated using a momentum update. Therefore, features can remain distinguishable and consistent, promoting inter-slice fusion, as shown in Fig.~\ref{fig:1} (b). Furthermore, we propose an efficient inter-slice fusion Transformer (IF-Trans) that captures inter-slice cues from multi-scale feature maps at each scale, built upon {\tt Swin Transformer}~\cite{liu2021swin}.

The main contributions of this work are summarized as follows:
\begin{itemize}
    \item A novel 2.5D-based model {\tt MOSformer} is proposed to fully exploit inter-slice information for 3D medical image segmentation.
    \item To make slice features distinguishable and consistent, dual encoders with a momentum update are introduced. Moreover, the inter-slice fusion transformer (IF-Trans) module is developed to efficiently fuse inter-slice information.
    \item State-of-the-art segmentation performance has been achieved by our model on three benchmark datasets, including Synapse, ACDC, and AMOS.
\end{itemize}

The remainder of this paper is organized as follows: Section~\ref{sec:s2} briefly reviews current segmentation methods. Section~\ref{sec:s3} depicts the proposed model in detail. Section~\ref{sec:s4} introduces model configurations and datasets. The experimental results are presented in Section~\ref{sec:s5}. Finally, Section~\ref{sec:s6} concludes this article.

\section{Related Works} \label{sec:s2}
\begin{figure*}[t]
\centering
\centerline{\includegraphics{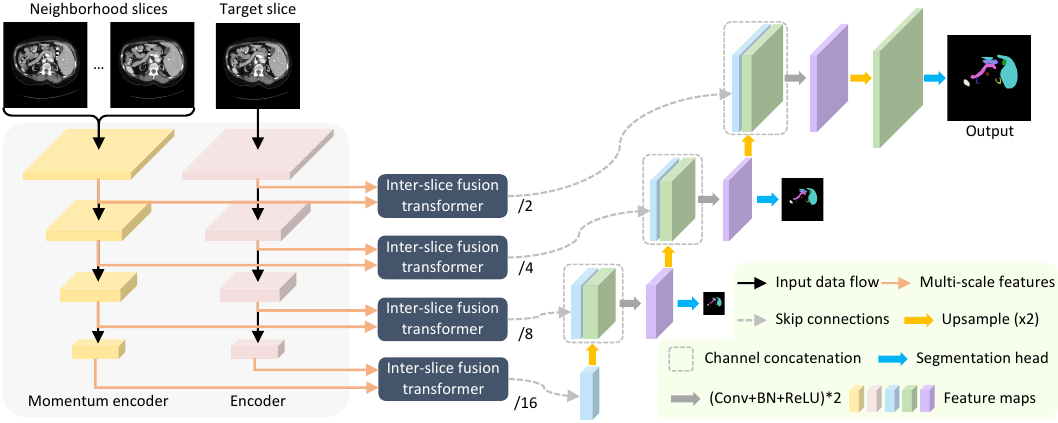}}
\caption{The architecture of {\tt MOSformer}. It comprises dual encoders: a momentum encoder that extracts features of the neighborhood slice and an encoder that extracts features from the target slice. IF-Trans is designed to perform inter-slice fusion independently at different scales. The fused features are then fed into a CNN decoder to produce segmentation maps for the target slices. Cylinders in yellow, pink, blue, green, and purple denote feature maps produced by the momentum encoder, the encoder, the IF-Trans, the upsampling operators, and the decoder blocks, \textit{i.e.},(Conv+BN+ReLU) * 2, respectively.}
\label{fig:2}
\end{figure*}
\subsection{2.5D-based Medical Image Segmentation}
Several 2.5D-based approaches have been proposed for efficient medical image segmentation by leveraging  inter-slice information. Early methods concatenated multiple consecutive 2D slices into a multi-channel input and adopted 2D-based models to segment specific regions of the middle slice~\cite{yu2018recurrent},~\cite{li2021acenet}. However, concatenating 2D slices as multi-channel inputs hinders the model’s capacity to disentangle and learn slice-specific features~\cite{zhang2022bridging}, thereby constraining the performance of 2.5D-based models. To solve the above problem, some works treated continuous 2D slices as temporal sequences and utilized recurrent neural networks (RNNs)~\cite{chen2016combine},~\cite{yang2018towards} to learn inter-slice information. For example, Chen \textit{et al.}~\cite{chen2016combine} introduced a 2.5D segmentation framework that combines $k$-UNet and bi-directional convolutional LSTM (BDC-LSTM) to integrate inter-slice information. Although RNNs can help improve the performance of 2.5D-based models to some extent, training costs of these methods are considerably high~\cite{pang2019deep}. Instead of RNNs, recent studies utilized attention mechanisms or transformers to fuse inter-slice information at the feature level effectively. Zhang \textit{et al.}~\cite{zhang2020sau} proposed an attention fusion module to refine segmentation results by fusing the information of adjacent slices. Li \textit{et al.}~\cite{li2021learning} employed a 2.5D coarse‑to‑fine architecture that leveraged inter‑slice prediction discrepancies as spatial attention cues to refine the initial segmentation. Guo \textit{et al.}~\cite{guo2021transformer} adopted 2D {\tt UNet} as the backbone and fused inter-slice information via a transformer at the bottom layer of the encoder. Yan \textit{et al.}~\cite{yan2022after} proposed {\tt AFTer-UNet} with an axial fusion mechanism based on transformer to fuse intra- and inter-slice contextual information. Hung \textit{et al.} \cite{hung2022cat},~\cite{hung2024csam} and Kumar \textit{et al.}~\cite{kumar2024flexible} introduced novel cross-slice attention mechanisms based on transformer to learn cross-slice information at multiple scales. However, the aforementioned methods fail to capture useful inter-slice information for the target slice which needs to be segmented, since they use a single encoder to extract slice features, making it difficult for models to distinguish target slices from neighborhood slices~\cite{zhang2022bridging}.

Different from previous works, we adopt dual encoders with a momentum update to extract features of target slices and neighborhood slices, respectively. We demonstrate that such a design can make features of target slices and neighborhood slices distinguishable and consistent, further boosting inter-slice fusion.

\subsection{Transformers in Medical Image Segmentation}
Recently, with the tremendous success of vision transformer (ViT)~\cite{dosovitskiy2020image} in various computer vision tasks~\cite{carion2020end},~\cite{lijin2024dual}, many works have explored using transformers in medical image segmentation. Compared with CNNs, transformers can capture long-range dependencies by sequence modeling and multihead self-attention (MHSA)~\cite{dosovitskiy2020image}, achieving better segmentation performance. Chen \textit{et al.}~\cite{chen2024transunet} proposed a hybrid model, {\tt TransUNet}, combining {\tt UNet}~\cite{ronneberger2015u} and transformer, where the transformer encodes feature maps from the CNN encoder to extract global contexts for the decoder to generate segmentation results. To fully unleash the transformer's potential, subsequent research focused on pure transformer architectures. A key challenge is the high computational complexity of self-attention on high-resolution medical images. {\tt Swin Transformer}~\cite{liu2021swin}, with its efficient window-based attention, offered a viable solution. Building on this, Cao \textit{et al.}~\cite{cao2022swin} introduced {\tt Swin-Unet}, the first pure transformer for medical image segmentation, which replaces all convolutions in {\tt U-Net} with {\tt Swin Transformer} blocks. However, this architecture does not achieve better performance than hybrid models \cite{you2022class}. Huang \textit{et al.} introduced {\tt MISSformer}~\cite{huang2023missformer}, which incorporates an encoder-decoder architecture built on enhanced transformer blocks. These blocks are connected through the ReMixed transformer context bridge, enhancing the model's ability to capture discriminative details. You \textit{et al.}~\cite{you2022class} presented {\tt CASTformer} with a class-aware transformer module to better capture discriminative regions of target objects. Moreover, they utilized adversarial learning to boost segmentation accuracies. However, the 2D-based methods mentioned above face limitations in leveraging inter-slice information, which hinders their potential for further performance improvements. Some attempts have been made to build 3D-based transformer segmentation models. {\tt UNETR}~\cite{hatamizadeh2022unetr} pioneered the use of a transformer-based encoder to learn global contexts from volumetric data. {\tt CoTr}~\cite{xie2021cotr} introduced a deformable self-attention mechanism to reduce computational complexity. However, simplifying self-attention may cause contextual information loss~\cite{yan2022after}. {\tt nnFormer}~\cite{zhou2023nnformer} is an interleaved architecture, where convolution layers encode precise spatial information and transformer layers fully explore global dependencies. Similar to {\tt Swin Transformer}~\cite{liu2021swin}, a computationally efficient way to calculate self-attention is proposed in {\tt nnFormer}.

In this work, we introduce the inter-slice fusion transformer (IF-Trans), which extends {\tt Swin Transformer}'s (shifted) window multi-head self-attention~\cite{liu2021swin}, (S)W-MSA, to fuse inter-slice information. Unlike prior methods that restrict transformer attention to each 2D slice, IF-Trans lifts (S)W-MSA into the inter-slice domain. Each window attends not only to patches within its own slice but also to the corresponding windows in neighborhood slices. This design preserves {\tt Swin Transformer}’s computational efficiency while jointly modeling intra-slice and inter-slice context, improving 3D medical image segmentation.

\section{Method} \label{sec:s3}
\subsection{Overall Architecture}
The detailed architecture of {\tt MOSformer} is shown in Fig.~\ref{fig:2}. Like most previous works for medical image segmentation, we utilize a hybrid encoder-decoder architecture, combining the advantages of CNNs and transformers~\cite{zhou2023nnformer}. $\bm{x}_i \in \mathbb{R}^{C\times H\times W}$ is the input of the encoder and represents the target slice for segmentation, where $i$ indicates the $i$-th slice of a 3D volume $\bm{X}\in \mathbb{R}^{C\times H\times W \times D}$, where $C$, $H$, $W$, and $D$ denote the channels, height, width, and depth of $\bm{X}$. $\bm{x}_j \in \mathbb{R}^{C\times H\times W}$ denotes an input to the momentum encoder corresponding to a neighborhood slice of $\bm{x}_i$, where $j\in \left[i-s, i+s\right]\setminus\left\{i\right\}$. The hyperparameter $s$ represents the $s$-th neighborhood of $\bm{x}_i$. Finally, the model generates the segmentation map $\bm{y}_i \in \mathbb{R}^{C_0\times H \times W}$ of $\bm{x}_i$, where $C_0$ is the number of label classes.

We adopt a lightly modified {\tt ResNet-50}~\cite{he2016deep} to implement the encoders, which extract multi-scale features from input slices. Concretely, the final downsampling stage is replaced with a non-downsampling stage to preserve spatial resolution. Dual encoders with a momentum update are adopted in {\tt MOSformer} to strengthen feature distinguishability and maintain feature consistency. Furthermore, IF-Trans modules are used at multi-scale ($1/2$, $1/4$, $1/8$, and $1/16$) to fuse inter-slice features extracted by dual encoders (details are provided in Section~\ref{sec:sec3_fusion}). Then the fused features are sent to the decoder via skip connections. The final segmentation predictions are derived via a segmentation head ($1\times 1$ convolutional layer).

\subsection{Dual Encoders with A Momentum Update}
Conventional 2.5D-based methods employ a single encoder to process both the target slice and its neighborhood slices, then fuse their features at a later stage. However, because all slices share the same feature space, the model struggles to distinguish target-specific cues from neighborhood context~\cite{zhang2022bridging}, as illustrated in Fig.~\ref{fig:1} (a). Consequently, inter-slice fusion may be suboptimal, and fine-grained spatial cues of the target slice can be lost. An intuitive solution is to use two separate encoders to process neighborhood slices and target slices, respectively. In practice, however, updating the parameters of these encoders independently during training leads to inconsistent feature distributions, which again hampers effective fusion.

\begin{figure*}[t]
\centering
\centerline{\includegraphics{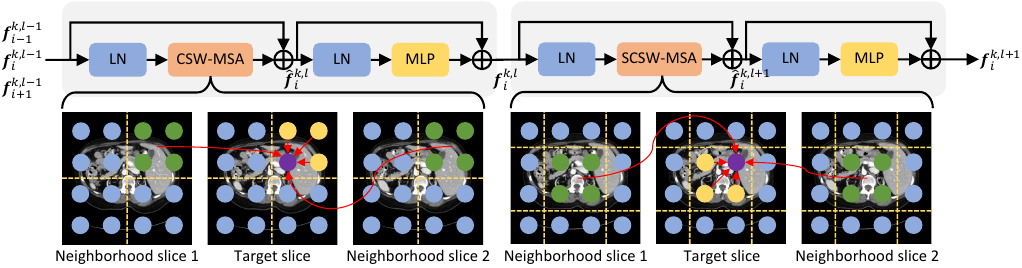}}
\caption{Schematic of inter-slice fusion transformer (IF-Trans) module. The neighborhood slice number is set to $1$ in this figure, consistent with our default model configuration. It has two successive IF-Trans with different window partitioning configurations. The colored circles indicate feature pixels. The window-based self-attention is expanded to the inter-slice dimension, promoting target slice feature pixels to learn intra- and inter-slice contexts. The black and red arrows denote the data flow within the IF-Trans Module and the fusion process of the IF-Trans Module.}
\label{fig:3}
\end{figure*}

To address this, we draw inspiration from the momentum contrast framework (MoCo)~\cite{he2020momentum} and introduce a momentum encoder for the neighborhood slices. The key idea is to maintain a slowly updating copy of the target encoder, ensuring that neighborhood features remain consistent over training while still being distinguishable from the target features.

Formally, let $\bm{\theta}_1$ denote the parameters of the target slice encoder, updated via standard back-propagation. We initialize the neighborhood slice encoder with parameters $\bm{\theta}_2 = \bm{\theta}_1$, and thereafter update it at each iteration as:
\begin{align}
	\bm{\theta}_2 \leftarrow m * \bm{\theta}_2 + (1-m) * \bm{\theta}_1 \label{eq:mom}
\end{align}
where $m \in \left[0, 1\right)$ is a momentum coefficient ($0.1$ by default). We analyze the impact of $m$ in Section~\ref{sec:s5_p3}, finding that a relatively small momentum yields the best results.

This momentum update strikes a balance between feature consistency and distinguishability. Specifically, neighborhood slice features are extracted by an encoder whose parameters update smoothly based on the target slice encoder, reducing abrupt shifts in feature space. At the same time, because $\bm{\theta}_2$ lags slightly behind $\bm{\theta}_1$, neighborhood features remain systematically distinct from target features, helping the fusion module to distinguish intra-slice and inter-slice information more effectively.

\subsection{Inter-slice Fusion Transformer}~\label{sec:sec3_fusion}
In this section, inter-slice fusion transformer (IF-Trans) is proposed to capture inter-slice cues, as shown in Fig.~\ref{fig:3}. We utilize IF-Trans at multiple scales and discuss the benefit of multi-scale learning in Section~\ref{sec:s5_p3}. Inputs of the $k$-th IF-Trans are feature maps $\left\{\bm{f}_{i-s}^k, \cdots, \bm{f}_i^k, \cdots, \bm{f}_{i+s}^k\right\}$ extracted by the encoder and the momentum encoder, where $k$ represents the $k$-th scale of two encoders ($k=1, 2, 3, 4$). The neighborhood slice number $s$ is set to $1$ in our default configuration. We give a detailed analysis of $s$ in Section~\ref{sec:s5_p3}. Therefore, the model uses adjacent ($1$-st neighborhood) slices of the target slice $\bm{x}_i$ as additional inputs.

Different from standard self-attention~\cite{Vaswani2017attn} with quadratic complexity, the proposed IF-Trans only calculates self-attention within the local window. As shown in the left part of Fig.~\ref{fig:3}, feature maps are partitioned into several non-overlapping windows\footnote{For intuitive explanation, feature maps are replaced by input images, and the number of feature pixels is simplified to 16.}. Compared with {\tt Swin Transformer}~\cite{liu2021swin}, we compute self-attention within inter- and intra-slice local windows (\textit{i.e.,} CSW-MSA, cross-slice window-based multi-head self-attention) instead of intra-slice local windows. Consequently, the purple feature pixel in the target slice attends not only to the yellow pixels within its own slice but also to the green pixels from neighborhood slices, thereby capturing both intra-slice and inter-slice context, as illustrated by the red arrows in Fig.~\ref{fig:3}.

However, the local CSW-MSA lacks connections across windows, reducing its representational power. Similar to~\cite{liu2021swin}, a shifted window partitioning strategy is introduced, allowing each pixel to receive broader views from intra- and inter-slices. In Fig.~\ref{fig:3}, the first transformer module adopts a regular window partition approach, and the feature map is evenly divided into $2\times 2$ windows of size $2\times 2$ ($M=2$)\footnote{To correspond with Fig.~\ref{fig:3}, $M$ is set to 2 here.}. The second transformer module uses a different partitioning configuration. Windows of the preceding layer are displaced by $\left(\lfloor\frac{M}{2}\rfloor, \lfloor\frac{M}{2}\rfloor\right)$ pixels to generate new windows. By doing so, the orange pixel can conduct self-attention (\textit{i.e.}, SCSW-MSA, a shifted version of CSW-MSA) with more pixels, thereby boosting its representational capacity. In practice, these two configurations are served as two consecutive layers to get an IF-Trans module. Outputs of IF-Trans can be formulated as:
\begin{flalign}
&\hat{\bm{f}}^{k, l}_i = \mathcal{T}\left\{\textit{\rm LN}\left(\bm{f}^{k, l-1}_{i-1}\right), \textit{\rm LN}\left(\bm{f}^{k, l-1}_i\right), \textit{\rm LN}\left(\bm{f}^{k, l-1}_{i+1}\right) \right\} + \bm{f}^{k, l-1}_i \notag && \\ 
&\bm{f}^{k, l}_i = \mathcal{M}\left\{\textit{\rm LN}\left(\hat{\bm{f}}^{k, l}_i\right)\right\} + \hat{\bm{f}}^{k, l}_i \notag && \\ 
& \hat{\bm{f}}^{k, l+1}_i = \mathcal{T}^{\rm S}\left\{\textit{\rm LN}\left(\bm{f}^{k, l}_{i-1}\right), \textit{\rm LN}\left(\bm{f}^{k, l}_i\right), \textit{\rm LN}\left(\bm{f}^{k, l}_{i+1}\right) \right\} + \bm{f}^{k, l}_i \notag && \\ 
&\bm{f}^{k, l+1}_i = \mathcal{M}\left\{\textit{\rm LN}\left(\hat{\bm{f}}^{k, l+1}_i\right)\right\} + \hat{\bm{f}}^{k, l+1}_i &&
\end{flalign}
where $\hat{\bm{f}}^{k, l}_i$ and $\bm{f}^{k,l}_i$ represents output feature maps of the (S)CSW-MSA module $\mathcal{T}^{\rm(S)}$ and the multilayer perceptron (MLP) module $\mathcal{M}$ in the $l$-th layer, respectively. LN indicates layer normalization. The query-key-value (QKV) self-attention~\cite{Vaswani2017attn} in (S)CSW-MSA is computed as follows:
\begin{align}
    \textit{\rm Attention}(\bm{Q},\bm{K},\bm{V})=\textit{\rm Softmax}\left(\frac{\bm{Q}\bm{K}^{\rm T}}{\sqrt{d}}+\bm{B}\right)\bm{V}
\end{align}
where $\bm{Q}\in \mathbb{R}^{\left\{M^2*\left(2*s+1\right)\right\}\times d}$, $\bm{K}\in \mathbb{R}^{\left\{M^2*\left(2*s+1\right)\right\}\times d}$, and $\bm{V}\in \mathbb{R}^{\left\{M^2*\left(2*s+1\right)\right\}\times d_0}$ denote query, key, and value matrices. $d$ and $d_0$ are embedding dimensions of query/key and value. In practice, $d$ is equal to $d_0$. $\bm{B}$ represents the position embedding matrix, and values are taken from the bias matrix $\hat{\bm{B}}\in \mathbb{R}^{(2M-1)\times (2M-1)}$.

\subsection{Loss Function}
Following previous methods~\cite{xie2021cotr,hatamizadeh2022unetr,zhou2023nnformer}, our model is trained end-to-end using the deep supervision strategy~\cite{lee2015deeply}. As illustrated in Fig.~\ref{fig:2}, final segmentation results are generated by the segmentation head ($1\times 1$ convolutional layer). Additionally, two smaller resolutions of decoder outputs are selected as auxiliary supervision signals. The deep supervision path in Fig.~\ref{fig:2} consists of an upsample layer and a $1\times 1$ convolutional layer. Therefore, the loss function can be formulated as follows:
\begin{align}
    \mathcal{L}_{\rm seg} = \lambda_1\mathcal{L}_{\left\{H,W\right\}} + \lambda_2\mathcal{L}_{\left\{\frac{H}{2},\frac{W}{2}\right\}} + \lambda_3\mathcal{L}_{\left\{\frac{H}{4},\frac{W}{4}\right\}}
\end{align}
where $\lambda_1$, $\lambda_2$, and $\lambda_3$ are $\frac{1}{2}$, $\frac{1}{4}$, and $\frac{1}{8}$, respectively. $\mathcal{L}_{\{h,w\}}$ represents the loss function on $h\times w$ resolution. It is a linear combination of cross-entropy loss $\mathcal{L}_{\rm CE}$ and Dice loss $\mathcal{L}_{\rm DSC}$:
\begin{align}
    \mathcal{L}_{\left\{h,w\right\}} = \alpha_1\mathcal{L}_{\rm CE} + \alpha_2 \mathcal{L}_{\rm DSC}
\end{align}
where $\alpha_1$ and $\alpha_2$ are 0.8 and 1.2, respectively.

\section{Experimental Setup} \label{sec:s4}
\subsection{Datasets}
To thoroughly compare {\tt MOSformer} to previous methods, we conduct experiments on three challenging benchmarks: the Synapse multi-organ segmentation dataset \cite{landman2015miccai}, the automated cardiac diagnosis challenge (ACDC) dataset \cite{bernard2018deep}, and the abdominal organ segmentation (AMOS) dataset \cite{ji2022amos}.

\textbf{Synapse for Multi-organ Segmentation.} This dataset consists of 30 abdominal CT scans with 8 organs (\emph{aorta}, \emph{gallbladder}, \emph{left kidney}, \emph{right kidney}, \emph{liver}, \emph{pancreas}, \emph{spleen}, and \emph{stomach}). Each volume has $85\sim 198$ slices of $512\times 512$ pixels. Following the splits adopted in {\tt TransUNet}~\cite{chen2024transunet}, the dataset is divided into 18 training cases and 12 testing cases.

\textbf{ACDC for Automated Cardiac Diagnosis Challenge.} The ACDC dataset includes cardiac MRI images of 100 patients from real clinical exams with manual annotations of \emph{left ventricle} (LV), \emph{right ventricle} (RV), and \emph{myocardium} (Myo). Consistent with {\tt TransUNet}~\cite{chen2024transunet}, the dataset is split into 70 training cases, 10 validation cases, and 20 testing cases.

\textbf{AMOS for Abdominal Organ Segmentation.} The AMOS dataset is a comprehensive abdominal organ segmentation dataset that includes patient annotations of 15 abdominal organs (\emph{aorta}, \emph{bladder}, \emph{duodenum}, \emph{esophagus}, \emph{gallbladder}, \emph{inferior vena cava}, \emph{left adrenal gland}, \emph{left kidney}, \emph{liver}, \emph{pancreas}, \emph{prostate/uterus}, \emph{right adrenal gland}, \emph{right kidney}, \emph{spleen}, and \emph{stomach}) from different centers, modalities, scanners, phases, and diseases. Only CT scans are utilized in our experiments, consisting of 200 training cases and $100$ testing cases.

\subsection{Implementation Details}
All experiments are implemented based on PyTorch 1.12.0, Python 3.8, and Ubuntu 18.04. Our model is trained on a single Nvidia A6000 GPU with 48GB of memory. The same model configurations are utilized on three datasets. Input medical images are resized to $224\times 224$ for a fair comparison. SGD optimizer with momentum of $0.9$ and weight decay of $1e^{-4}$ is adopted to train our model for 300 epochs. The batch size is set to 24. A cosine learning rate scheduler with five epochs of linear warm-up is used during training, and the maximum and minimum learning rates are $3e^{-2}$ and $5e^{-3}$, respectively.

\subsection{Evaluation Metrics}
Two metrics are utilized to evaluate segmentation performance of models: Dice similarity score (DSC), and 95\% Hausdorff distance (HD95).

DSC is utilized to evaluate overlaps between ground truths and segmentation results and is defined as follows:
\begin{align}
    &{\rm DSC}(P,G) = 2\times\frac{|P\cap G|}{|P|+|G|}
\end{align}
where $P$ refers to model predictions and $G$ refers to ground truths.

HD95 is adopted to measure the 95\% distance between boundaries of model predictions and ground truths. It is defined as follows:
\begin{align}
    {\rm HD}_{95}=\max\left\{d_{PG},d_{GP}\right\}
\end{align}
where $d_{PG}$ is the maximum 95\% distance between model predictions and ground truths. $d_{GP}$ is the maximum 95\% distance between ground truths and model predictions.

\section{Results} \label{sec:s5}
\subsection{Comparisons with SOTAs}
\begin{table*}[t]
\caption{Comparison with state-of-the-art models on the multi-organ segmentation (Synapse) dataset. The best results are highlighted in \textbf{bold} and the second-best results are \underline{underlined}. The evaluation metrics are DSC and HD95, consistent with TransUNet~\cite{chen2024transunet}. Moreover, DSC of each organ is reported in this table. $^\ddagger$ and $^\dagger$ indicate the results are borrowed from~\cite{zhou2023nnformer} and~\cite{cao2022swin}, respectively. $^\ast$ means the baselines are implemented by ourselves. Baselines without any symbol represent the results are from the original papers.}
\label{table:synapse}
\centering
\renewcommand\arraystretch{1.2}
\resizebox{\linewidth}{!}{\begin{tabular}{llcccccccccc}
\toprule
\multirow{2}{*}{Dimension} & \multirow{2}{*}{Method} & \multicolumn{2}{c}{Average} & \multirow{2}{*}{Aorta} & \multirow{2}{*}{Gallbladder} & \multirow{2}{*}{Kidney (L)} & \multirow{2}{*}{Kidney (R)} & \multirow{2}{*}{Liver} & \multirow{2}{*}{Pancreas} & \multirow{2}{*}{Spleen} & \multirow{2}{*}{Stomach} \\ \cmidrule{3-4}
 & & DSC (\%) $\uparrow$ & HD95 (mm) $\downarrow$ & & & & & & & & \\ \midrule
\multirow{11}{*}{2D} & UNet$^\dagger$~\cite{ronneberger2015u} {\tiny \color{gray} [MICCAI'15]} & 76.85 & 39.70 & 89.07 & 69.72 & 77.77 & 68.60 & 93.43 & 53.98 & 86.67 & 75.58 \\ 
& AttnUNet$^\dagger$ \cite{schlemper2019attention} {\tiny \color{gray} [MedIA'19]} & 77.77 & 36.02 & 89.55 & 68.88 & 77.98 & 71.11 & 93.57 & 58.04 & 87.30 & 75.75 \\
& TransUNet \cite{chen2024transunet} {\tiny \color{gray} [MedIA'24]} & 77.48 & 31.69 & 87.23 & 63.13 & 81.87 & 77.02 & 94.08 & 55.86 & 85.08 & 75.62 \\
& MISSFormer \cite{huang2023missformer} {\tiny \color{gray} [TMI'23]} & 81.96 & 18.20 & 86.99 & 68.65 & 85.21 & 82.00 & 94.41 & 65.67 & 91.92 & 80.81 \\
& SwinUNet \cite{cao2022swin} {\tiny \color{gray} [ECCVW'22]} & 79.12 & 21.55 & 85.47 & 66.53 & 83.28 & 79.61 & 94.29 & 56.58 & 90.66 & 76.60 \\
& MT-UNet \cite{wang2022mixed} {\tiny \color{gray} [ICASSP'22]} & 78.59 & 26.59 & 87.92 & 64.99 & 81.47 & 77.29 & 93.06 & 59.46 & 87.75 & 76.81 \\
& UCTransNet \cite{wang2022uctransnet} {\tiny \color{gray} [AAAI'22]} & 78.23 & 26.75 & 88.86 & 66.97 & 80.19 & 73.18 & 93.17 & 56.22 & 87.84 & 79.43 \\
& CASTformer \cite{you2022class} {\tiny \color{gray} [NeurIPS'22]} & 82.55 & 22.73 & 89.05 & 67.48 & 86.05 & 82.17 & 95.61 & 67.49 & 91.00 & 81.55 \\
& HiFormer \cite{heidari2023hiformer} {\tiny \color{gray} [WACV'23]} & 80.39 & 14.70 & 86.21 & 65.69 & 85.23 & 79.77 & 94.61 & 59.52 & 90.99 & 81.08 \\
& SAMed \cite{zhang2023customized} {\tiny \color{gray} [arXiv'23]} & 81.88 & 20.64 & 87.77 & 69.11 & 80.45 & 79.95 & 94.80 & 72.17 & 88.72 & 82.06 \\
\midrule
\multirow{6}{*}{3D} & V-Net$^\dagger$ \cite{milletari2016v} {\tiny \color{gray} [3DV'16]} & 68.81 & - & 75.34 & 51.87 & 77.10 & 80.75 & 87.84 & 40.05 & 80.56 & 56.98 \\
& CoTr$^\ddagger$ \cite{xie2021cotr} {\tiny \color{gray} [MICCAI'21]} & 80.78 & 19.15 & 85.42 & 68.93 & 85.45 & 83.62 & 93.89 & 63.77 & 88.58 & 76.23 \\
& UNETR$^\ddagger$ \cite{hatamizadeh2022unetr} {\tiny \color{gray} [WACV'22]} & 79.56 & 22.97 & 89.99 & 60.56 & 85.66 & 84.80 & 94.46 & 59.25 & 87.81 & 73.99 \\
& SwinUNETR$^\ddagger$ \cite{hatamizadeh2021swin} {\tiny \color{gray} [MICCAIW'22]} & 83.51 & 14.78 & 90.75 & 66.72 & 86.51 & \underline{85.88} & 95.33 & 70.07 & \textbf{94.59} & 78.20 \\
& nnFormer \cite{zhou2023nnformer} {\tiny \color{gray} [TIP'23]} & \textbf{86.57} & \textbf{10.63} & \textbf{92.04} & 70.17 & 86.57 & \textbf{86.25} & \textbf{96.84} & \textbf{83.35} & 90.51 & \underline{86.83} \\
& SAM3D \cite{bui2024sam3d} {\tiny \color{gray} [ISBI'24]} & 79.56 & 17.87 & 89.57 & 49.81 & 86.31 & 85.64 & 95.42 & 69.32 & 84.29 & 76.11 \\
\midrule
\multirow{4}{*}{2.5D} & AFTer-UNet \cite{yan2022after} {\tiny \color{gray} [WACV'22]} & 81.02 & - & \underline{90.91} & 64.81 & \underline{87.90} & 85.30 & 92.20 & 63.54 & 90.99 & 72.48 \\
& TransUNet-2.5D \cite{zhang2023multi} {\tiny \color{gray} [TIM'23]} & 84.24 & 19.24 & 89.97 & \textbf{73.28} & 83.99 & 81.33 & 95.61 & 70.39 & \textbf{94.59} & 84.78 \\
& CSA-Net$^\ast$ \cite{kumar2024flexible} {\tiny \color{gray} [CIBM'24]} & 79.96 & 32.11 & 83.91 & 64.99 & 83.56 & 79.93 & 94.43 & 62.65 & 91.12 & 79.12 \\
& \textbf{MOSformer} {\tiny \color{gray} \textbf{[Ours]}} & \underline{85.63} & \underline{13.40} & 88.95 & \underline{71.90} & \textbf{90.32} & 83.58 & \underline{95.96} & \underline{74.14} & \underline{92.29} & \textbf{87.87} \\
\bottomrule
\end{tabular}}

\end{table*}

We select several state-of-the-art 2D, 3D, and 2.5D medical image segmentation models as our baselines. To ensure a fair comparison, all models are trained and evaluated using identical preprocessing pipelines. Specifically, we apply the {\tt TransUNet} preprocessing protocol~\cite{chen2024transunet} to both the multi-organ segmentation (Synapse) and the automated cardiac diagnosis challenge (ACDC) datasets, and follow the preprocessing procedures of~\cite{ji2022amos} for the abdominal organ segmentation (AMOS) dataset. It should be noted that we only visualize selected qualitative results from some representative models for clarity and visual impact.

\textbf{Multi-organ Segmentation (Synapse).} Quantitative results of state-of-the-art models and our {\tt MOSformer} are presented in Table~\ref{table:synapse}. {\tt MOSformer} achieves $85.63\%$ DSC and $13.40$ ${\rm mm}$ HD95 on this dataset. Compared with the best 2D-based method, \textit{i.e.,} {\tt CASTformer}~\cite{you2022class}, {\tt MOSformer} is able to surpass it by a large margin ($+3.08\%$ DSC and $-9.33$ ${\rm mm}$ HD95). For 2.5D-based baselines, {\tt MOSformer} demonstrates notable performance enhancements, offering at least $+4.61\%$, $+1.39\%$, and $+5.67\%$ DSC gains over {\tt AFTer-UNet}~\cite{yan2022after}, {\tt TransUNet-2.5D}~\cite{zhang2023multi}, and {\tt CSA-Net}~\cite{kumar2024flexible}, respectively. These results indicate \textbf{i)} the necessity of inter-slice information in 3D medical image segmentation; and \textbf{ii)} the effectiveness of distinguishable and consistent slice features produced by dual encoders with a momentum update.

\begin{figure*}[t]
\centering
\centerline{\includegraphics{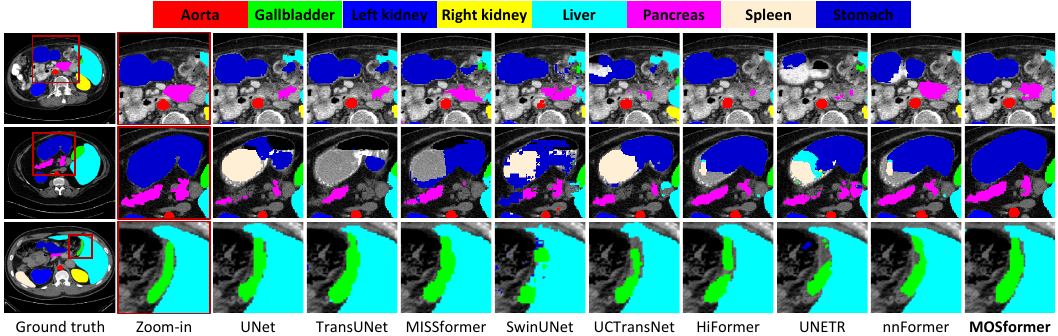}}
\caption{Visual comparisons with some representative methods on the multi-organ segmentation (Synapse) dataset.}
\label{fig:4}
\end{figure*}

\begin{table}[htbp]
\caption{Comparison with the state-of-the-art models on the automated cardiac diagnosis challenge (ACDC) dataset. The best results are highlighted in \textbf{bold} and the second-best results are \underline{underlined}. We only report DSC in this table, following the evaluation setting of TransUNet~\cite{chen2024transunet}. Moreover, DSC of each anatomical structure is reported in this table. $^\ddagger$ and $^\dagger$ indicate the results are borrowed from~\cite{zhou2023nnformer} and~\cite{wang2022mixed}, respectively. $^\ast$ means the baselines are implemented by ourselves. Baselines without any symbol represent the results are from the original papers.
}
\label{table:acdc}
\centering
\renewcommand\arraystretch{1.2}
\renewcommand\arraystretch{1.2}
\resizebox{\linewidth}{!}{
\begin{tabular}{llccccc}
\toprule
Dimension & Method & DSC (\%) $\uparrow$ & RV & Myo & LV \\ \midrule
\multirow{8}{*}{2D} & UNet$^\dagger$~\cite{ronneberger2015u} {\tiny \color{gray} [MICCAI'15]} & 87.60 & 84.62 & 84.52 & 93.68 \\ 
& AttnUNet$^\dagger$ \cite{schlemper2019attention} {\tiny \color{gray} [MedIA'19]} & 86.90 & 83.27 & 84.33 & 93.53 \\
& TransUNet \cite{chen2024transunet} {\tiny \color{gray} [MedIA'24]} & 89.71 & 86.67 & 87.27 & 95.18 \\
& MISSFormer \cite{huang2023missformer} {\tiny \color{gray} [TMI'23]} & 91.19 & 89.85 & 88.38 & 95.34 \\
& SwinUNet \cite{cao2022swin} {\tiny \color{gray} [ECCVW'22]} & 88.07 & 85.77 & 84.42 & 94.03 \\
& MT-UNet \cite{wang2022mixed} {\tiny \color{gray} [ICASSP'22]} & 90.43 & 86.64 & 89.04 & 95.62 \\
& UCTransNet$^\ast$ \cite{wang2022uctransnet} {\tiny \color{gray} [AAAI'22]} & 91.98 & 90.06 & \textbf{89.87} & \underline{96.02} \\
& HiFormer$^\ast$ \cite{heidari2023hiformer} {\tiny \color{gray} [WACV'23]} & 90.40 & 88.24 & 87.63 & 95.30 \\
\midrule
\multirow{3}{*}{3D} & UNETR$^\ddagger$ \cite{hatamizadeh2022unetr} {\tiny \color{gray} [WACV'22]} & 88.61 & 85.29 & 86.52 & 94.02 \\
& nnFormer \cite{zhou2023nnformer} {\tiny \color{gray} [TIP'23]} & \underline{92.06} & \textbf{90.94} & 89.58 & 95.65 \\
& SAM3D \cite{bui2024sam3d} {\tiny \color{gray} [ISBI'24]} & 90.41 & 89.44 & 87.12 & 94.67 \\
\midrule
\multirow{3}{*}{2.5D} & CAT-Net$^\ast$ \cite{hung2022cat} {\tiny \color{gray} [TMI'22]} & 90.02 & 86.05 & 88.75 & 95.27 \\
& CSA-Net$^\ast$ \cite{kumar2024flexible} {\tiny \color{gray} [CIBM'24]} & 89.58 & 86.56 & 86.91 & 95.26 \\
& \textbf{MOSformer} {\tiny \color{gray} \textbf{[Ours]}} & \textbf{92.19} & \underline{90.86} & \underline{89.65} & \textbf{96.05} \\
\bottomrule
\end{tabular}
}
\end{table}
\begin{figure}[htbp]
\centering
\centerline{\includegraphics{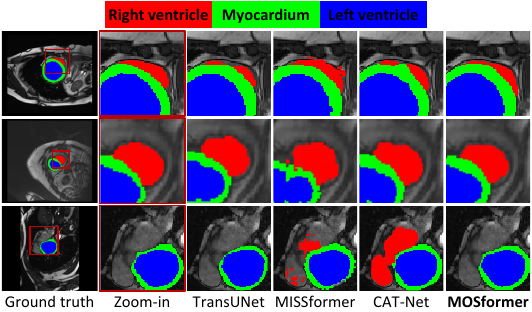}}
\caption{Visual comparisons with some representative methods on the automatic cardiac diagnosis challenge (ACDC) dataset.}
\label{fig:5}
\end{figure}

We also compare our {\tt MOSformer} with 3D-based segmentation methods. It still has competitive performance, surpassing five of the most widely recognized models and achieving comparable performance to {\tt nnFormer}~\cite{zhou2023nnformer}. It should be noted that {\tt MOSformer} obtains better DSC than {\tt nnFormer} in four organs (half of the categories), including \textit{gallbladder} ($+1.73\%$), \textit{left kidney} ($+3.75\%$), \textit{spleen} ($+1.78\%$), and \textit{stomach} ($+1.04\%$). Among these organs, \textit{gallbladder} and \textit{stomach} are two of the most difficult organs to segment since the \textit{gallbladder} is very small and the boundaries between the \textit{gallbladder} and the \textit{liver} are blurred while the \textit{stomach} has a significant intra-class variance. This reveals that our {\tt MOSformer} can learn more discriminative features and has a comprehensive understanding of organ structures.

Fig.~\ref{fig:4} shows qualitative comparisons of {\tt MOSformer} against several models on representative examples on the Synapse dataset. Most baselines suffer from segmentation target incompleteness (\textit{e.g.,} \textit{stomach}), misclassification of organs (\textit{e.g.,} \textit{spleen}), and blurry category boundaries (\textit{e.g., gallbladder}), while {\tt MOSformer} can locate organs precisely, reduce the number of false positive predictions, and produce sharper boundaries.

\textbf{Automated Cardiac Diagnosis Challenge (ACDC).} To further prove the model's generalization performance, {\tt MOSformer} is evaluated on the automated cardiac diagnosis challenge (ACDC) dataset. It should be noted that MRI images in this dataset can be considered anisotropic since they have high in-plane image resolution (\textit{e.g.,} $1.37\sim 1.68$ ${\rm mm}$) and low through-plane resolution (\textit{e.g.,} $5$ ${\rm mm}$)~\cite{bernard2018deep}. Quantitative results are summarized in Table~\ref{table:acdc}. Compared with state-of-the-art methods (2D, 2.5D, and 3D-based), {\tt MOSformer} achieves the best performance with $92.19\%$ DSC. Thus, the above results indicate that our 2.5D-based {\tt MOSformer} is more effective at processing anisotropic data compared with 3D-based models. Fig.~\ref{fig:5} presents qualitative comparisons for different methods on this dataset. As seen, {\tt MOSformer} can locate anatomical structures more accurately. Specifically, in case 3, many models mistakenly classify regions outside the \textit{myocardium} into the \textit{right ventricle} while {\tt MOSformer} does not produce any false positive predictions.

\begin{table*}[t]
\caption{Comparison with the state-of-the-art models on the abdominal organ segmentation (AMOS) dataset. The best results are highlighted in \textbf{bold} and the second-best results are \underline{underlined}. DSC is utilized as evaluation metric. Moreover, DSC of each organ is reported in this table. $^\ast$ means the baselines are implemented by ourselves.}
\label{table:amos}
\centering
\renewcommand\arraystretch{1.2}
\resizebox{\linewidth}{!}{\begin{tabular}{llcccccccccccccccc}
\toprule
Dimension & Method & DSC (\%) $\uparrow$ & Spleen & Kid. (R) & Kid. (L) & Gall. & Eso. & Liver & Stom. & Aorta & IVC & Panc. & Adr. (R) & Adr. (L) & Duo. & Blad. & Pros. \\ \midrule
\multirow{5}{*}{2D} & UNet$^\ast$ \cite{ronneberger2015u} {\tiny \color{gray} [MICCAI'15]} & \underline{82.53} & 92.25 & 92.45 & 92.50 & \textbf{81.85} & \underline{79.98} & 94.73 & 84.80 & \underline{92.20} & 82.94 & 77.35 & 67.13 & 69.34 & \underline{72.77} & 82.40 & 75.31 \\
& TransUNet$^\ast$ \cite{chen2024transunet} {\tiny \color{gray} [MedIA'24]} & 80.10 & 91.26 & 92.47 & 91.90 & 78.01 & 77.00 & 94.93 & 80.04 & 91.98 & 82.99 & 74.30 & 63.66 & 53.84 & 71.65 & 81.37 & 76.03 \\
& MISSFormer$^\ast$ \cite{huang2023missformer} {\tiny \color{gray} [TMI'23]} & 78.16 & 93.13 & 91.98 & 91.88 & 75.89 & 71.87 & 94.27 & 80.14 & 88.74 & 77.53 & 71.39 & 60.65 & 59.32 & 64.43 & 77.97 & 73.16 \\
& UCTransNet$^\ast$ \cite{wang2022uctransnet} {\tiny \color{gray} [AAAI'22]} & 82.34 & 93.37 & 92.32 & 91.90 & 77.09 & 79.77 & 94.78 & \underline{85.95} & 91.77 & 82.84 & \underline{77.44} & 65.88 & 68.98 & 71.36 & \underline{83.93} & \underline{77.71} \\
& HiFormer$^\ast$ \cite{heidari2023hiformer} {\tiny \color{gray} [WACV'23]} & 80.03 & 92.73 & 92.79 & 92.01 & 79.44 & 76.42 & 94.55 & 82.65 & 90.56 & 80.16 & 73.59 & 61.14 & 58.73 & 68.12 & 82.01 & 75.64 \\
\midrule
\multirow{2}{*}{3D} & UNETR$^\ast$ \cite{hatamizadeh2022unetr} {\tiny \color{gray} [WACV'22]} & 78.07 & \underline{93.38} & 93.00 & 92.28 & 73.17 & 69.72 & 94.86 & 73.25 & 90.82 & 80.20 & 73.44 & 65.19 & 60.69 & 65.46 & 74.10 & 71.49 \\
& nnFormer$^\ast$ \cite{zhou2023nnformer} {\tiny \color{gray} [TIP'23]} & 78.66 & 91.43 & 92.39 & 92.08 & 76.74 & 69.16 & 94.95 & 84.84 & 89.53 & 82.06 & 75.91 & 62.56 & 60.36 & 68.50 & 74.74 & 64.61 \\
\midrule
\multirow{2}{*}{2.5D} & CSA-Net$^\ast$ \cite{kumar2024flexible} {\tiny \color{gray} [CIBM'24]} & 82.12 & 91.25 & \underline{93.51} & \underline{93.68} & 79.01 & 78.80 & \underline{95.32} & 82.14 & 91.64 & \underline{83.94} & 75.18 & \underline{68.27} & \underline{69.37} & 71.36 & 83.00 & 75.33 \\
& \textbf{MOSformer} \textbf{{\tiny \color{gray} [Ours]}} & \textbf{85.43} & \textbf{95.26} & \textbf{94.68} & \textbf{94.54} & \underline{81.53} & \textbf{82.05} & \textbf{96.55} & \textbf{89.07} & \textbf{92.81} & \textbf{86.16} & \textbf{80.28} & \textbf{73.28} & \textbf{73.19} & \textbf{75.05} & \textbf{86.92} & \textbf{80.05} \\ \bottomrule
\end{tabular}}

\end{table*}
\begin{figure*}[t]
\centering
\centerline{\includegraphics{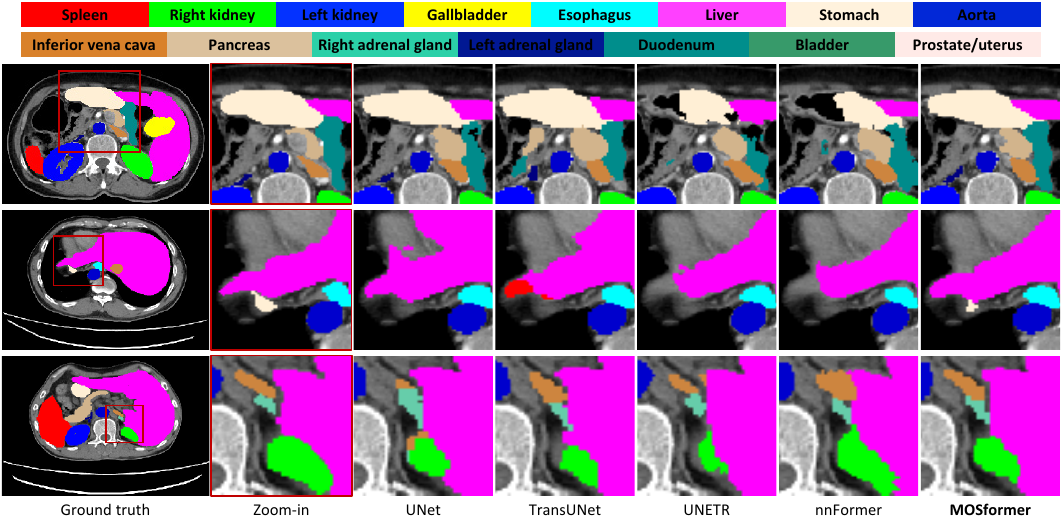}}
\caption{{Visual comparisons with some representative methods on the abdominal organ segmentation (AMOS) dataset.}}
\label{fig:6}
\end{figure*}

\textbf{Abdominal Organ Segmentation (AMOS).} Additionally, a large dataset with 200 training cases and 100 testing cases is also adopted in our experiments. Overall results and individual DSC on 15 organs are reported, as shown in Table~\ref{table:amos}. Our {\tt MOSformer} achieves the best DSC in 14 organs and the second-best DSC in one organ. Surprisingly, {\tt MOSformer} offers $+6.77\%$ DSC improvement over 3D-based {\tt nnFormer} while they have similar performance on the multi-organ segmentation (Synapse) dataset. Based on the above observation, it can be concluded that the performance of {\tt MOSformer} is more stable across different datasets compared with {\tt nnFormer}. Visualization results are shown in Fig.~\ref{fig:6}. Compared with baselines, our {\tt MOSformer} is able to accurately segment organs of diverse shapes and sizes, thus providing more consistent results with ground truths.

\textbf{Statistical Analysis.} In Table~\ref{table:pvalue}, we report Wilcoxon-test $p$-values on case-level DSC for paired comparisons between {\tt MOSformer} and the strongest publicly available 2D, 3D, and 2.5D baselines. On the Synapse dataset, although the 3D {\tt nnFormer} attains a slightly higher mean DSC than {\tt MOSformer}, the difference is not statistically significant ($p$ = 0.339), indicating comparable performance. In contrast, {\tt MOSformer} significantly outperforms the 2D {\tt MISSFormer} ($p$ < 0.001) and the 2.5D {\tt CSA-Net} ($p$ < 0.001). On the ACDC dataset, {\tt MOSformer} achieves the highest mean DSC, but the margins over {\tt UCTransNet} ($p$ = 0.745) and {\tt nnFormer} ($p$ = 0.826) are not significant, while the improvement over {\tt CAT-Net} is significant ($p$ < 0.001). On the AMOS dataset, {\tt MOSformer} shows significant gains over all baselines (all $p$ < 0.001), demonstrating consistent advantages. Notably, the AMOS dataset has a much larger test set ($N$ = 100) than the Synapse ($N$ = 12) and the ACDC ($N$ = 40) datasets, providing greater statistical power. Accordingly, $p$-values on the AMOS dataset are more sensitive to performance differences, whereas non-significant results on the Synapse or the ACDC datasets may reflect limited sample sizes. Overall, this analysis strengthens the empirical evidence for {\tt MOSformer}’s effectiveness across diverse datasets.

\begin{table*}[t]
\caption{
 Statistical analysis on the multi-organ segmentation (Synapse), automated cardiac diagnosis challenge (ACDC), and abdominal organ segmentation (AMOS) datasets. $N$ is the number of testing cases.
}
\label{table:pvalue}
\centering
\renewcommand\arraystretch{1.2}
\resizebox{\linewidth}{!}{\begin{tabular}{lcc|lcc|lcc}
\toprule
\multicolumn{3}{c|}{Synapse ($N$ = 12)} & \multicolumn{3}{c|}{ACDC ($N$ = 40)} & \multicolumn{3}{c}{AMOS ($N$ = 100)} \\ \midrule
Method & DSC & $p$ & Method & DSC & $p$ & Method & DSC & $p$ \\ \hline
MISSFormer \cite{huang2023missformer} {\tiny \color{gray} [TMI'23]} & 81.96 & < 0.001 & UCTransNet \cite{wang2022uctransnet} {\tiny \color{gray} [AAAI'22]} & 91.89 & 0.745 & UNet~\cite{ronneberger2015u} {\tiny \color{gray} [MICCAI'15]} & 82.53 & < 0.001 \\
nnFormer \cite{zhou2023nnformer} {\tiny \color{gray} [TIP'23]} & 86.57 & 0.339 & nnFormer \cite{zhou2023nnformer} {\tiny \color{gray} [TIP'23]} & 92.06 & 0.826 & nnFormer \cite{zhou2023nnformer} {\tiny \color{gray} [TIP'23]} & 78.66 & < 0.001 \\
CSA-Net \cite{kumar2024flexible} {\tiny \color{gray} [CIBM'24]} & 79.96 & < 0.001 & CAT-Net \cite{hung2022cat} {\tiny \color{gray} [TMI'22]} & 90.02 & < 0.001 & CSA-Net \cite{kumar2024flexible} {\tiny \color{gray} [CIBM'24]} & 82.12 & < 0.001 \\
\textbf{MOSformer} {\tiny \color{gray} \textbf{[Ours]}} & 85.63 & - & \textbf{MOSformer} {\tiny \color{gray} \textbf{[Ours]}} & 92.19 & - & \textbf{MOSformer} {\tiny \color{gray} \textbf{[Ours]}} & 85.43 & - \\
\bottomrule
\end{tabular}}

\end{table*}
\begin{table}[htbp]
\caption{Ablation study of each component on the multi-organ segmentation (Synapse), automated cardiac diagnosis challenge (ACDC), and abdominal organ segmentation (AMOS) datasets. Enc-S: Single encoder; Enc-D: Dual encoders; Enc-DM: Dual encoders with a momentum update. The best results are highlighted in \textbf{bold}. $^\dagger$ means the model is 2D-based.
}
\label{table:ablation}
\centering
\renewcommand\arraystretch{1.2}
\renewcommand\arraystretch{1.2}
\resizebox{\linewidth}{!}{
\begin{tabular}{lcccc|l|l|l}
\toprule
\multirow{2}{*}{Model} & \multicolumn{4}{c|}{Module} & Synapse & ACDC & AMOS \\ \cmidrule{2-8} 
& Enc-S & Enc-D & Enc-DM & IF-Trans & DSC (\%) $\uparrow$ & DSC (\%) $\uparrow$ & DSC (\%) $\uparrow$ \\ \midrule
Model-1$^\dagger$ & \Checkmark & & & & 82.42 {\color{mydarkgreen}(-3.21)} & 91.61 {\color{mydarkgreen}(-0.58)} & 81.28 {\color{mydarkgreen}(-4.15)} \\
Model-2 & \Checkmark & & & \Checkmark & 84.23 {\color{mydarkgreen}(-1.40)} & 92.04 {\color{mydarkgreen}(-0.15)} & 82.63 {\color{mydarkgreen}(-2.80)} \\
Model-3 & & \Checkmark & & \Checkmark & 84.93 {\color{mydarkgreen}(-0.70)} & 92.10 {\color{mydarkgreen}(-0.09)} & 83.88 {\color{mydarkgreen}(-1.55)} \\
\textbf{MOSformer} & & & \Checkmark & \Checkmark & \textbf{85.63} & \textbf{92.19} & \textbf{85.43} \\ \bottomrule
\end{tabular}
}
\end{table}
\begin{figure*}[t]
\centering
\centerline{\includegraphics{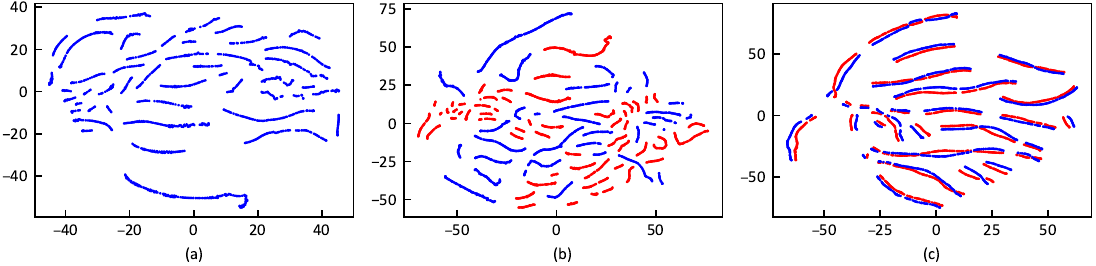}}
\caption{Visualization of embedding space learned under three encoder settings on the multi-organ segmentation (Synapse) \textit{test} set (1,568 slices). Each point represents the feature of a slice. Distinct colors are used to differentiate embeddings from different encoders. Because Model-2 uses a single encoder for both target and neighborhood slices, their embeddings are identical. Therefore, only the blue points are shown. Dimensions are reduced by t-SNE~\cite{van2008visualizing}. (a) Model-2 (Single encoder); (b) Model-3 (Dual encoders updated independently); (c) MOSformer (Dual encoders with a momentum update).}
\label{fig:tsne_vis}
\end{figure*}

\subsection{Ablation Study} \label{sec:s5_p2}
Extensive ablation studies are conducted on the multi-organ segmentation (Synapse), the automated cardiac diagnosis challenge (ACDC) and the abdominal organ segmentation (AMOS) datasets to verify the effectiveness of the momentum encoder and IF-Trans. DSC is selected as the default evaluation metric. Quantitative results are shown in Table~\ref{table:ablation}. It should be noted that the baseline, {\tt Model-1}, is a 2D-based model.

\textbf{Importance of The Momentum Update.} Two variants of {\tt MOSformer} are employed in this experiment: \textbf{i)} {\tt Model-2}: the encoder with a momentum update is removed, using a single encoder to extract features of target and neighborhood slices; \textbf{ii)} {\tt Model-3}: the momentum encoder is replaced by a normal encoder and parameters of two encoders are updated independently via back-propagation. From quantitative results presented in Table~\ref{table:ablation}, we can observe that these variants lead to decreased performance on the Synapse dataset ($+1.40\%$ and $+0.70\%$ in DSC), the ACDC dataset ($+0.15\%$ and $+0.09\%$ in DSC), and the AMOS dataset ($+2.80\%$ and $+1.55\%$ in DSC). The above results confirm the importance of the momentum update, designed to make slice features distinguishable and consistent. This design enables the model to distinguish target slices and fuse inter-slice information effectively.

Furthermore, we also adopt t-SNE~\cite{van2008visualizing} to visualize the encoded embedding space learned from three encoder settings on the multi-organ segmentation (Synapse) \textit{test} set. {\tt Model-2} employs a single encoder to process both target and neighborhood slices. Consequently, target and neighborhood slice features originate from the same feature space, as depicted in Fig.~\ref{fig:tsne_vis} (a). This setup poses challenges for the model in distinguishing individual slices and acquiring slice-specific information during inter-slice fusion. In contrast, the embedding space learned by dual encoders is distinguishable, as illustrated in Fig.~\ref{fig:tsne_vis} (b) and (c). It can also be observed that incorporating the momentum update in dual encoders facilitates consistency among slice features, as shown in Fig.~\ref{fig:tsne_vis} (c), thereby further boosting segmentation performance.

\begin{figure}[htbp]
\centering
\centerline{\includegraphics{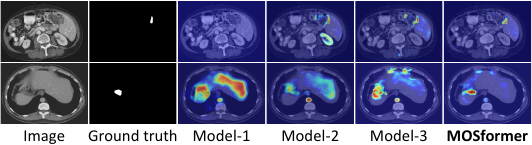}}
\caption{Class activation maps of the \textit{gallbladder} and the \textit{stomach} categories (from top to bottom) produced by Grad-CAM~\cite{selvaraju2017grad}. The class activation maps are generated from the last decoder layer.}
\label{fig:grad_cam}
\end{figure}
\begin{figure}[htbp]
\centering
\centerline{\includegraphics{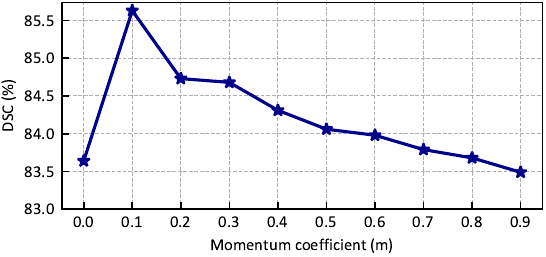}}
\caption{Effect of momentum coefficient $m$. We report DSC of {\tt MOSformer} on the multi-organ segmentation (Synapse) dataset.}
\label{fig:mom_coef}
\end{figure}

\textbf{Efficacy of The Inter-slice Fusion Transformer.} Compared to the baseline {\tt Model-1}, {\tt Model-2} with the IF-Trans module offers substantial improvements, increasing DSC by $+1.81\%$, $+0.43\%$, and $+1.35\%$ on the Synapse, ACDC, and AMOS datasets, respectively. Furthermore, enhancing feature discriminability with dual encoders yields even greater performance gains. Specifically, {\tt Model-3} and our {\tt MOSformer} further boost DSC by $+2.51\%$ and $+3.21\%$ on Synapse, $+0.49\%$ and $+0.58\%$ on ACDC, and $+2.60\%$ and $+4.15\%$ on AMOS, respectively.


Additionally, we employ Grad-CAM~\cite{selvaraju2017grad} to visualize discriminative regions of the models, as depicted in Fig.~\ref{fig:grad_cam}. Compared with baseline {\tt Model-1}, we can see that inter-slice information is beneficial, but {\tt Model-2} and {\tt Model-3} still tend to assign weights to irrelevant regions. Distinguishable and consistent inter-slice features within {\tt MOSformer} can address the above issue, demonstrating enhanced precision in localizing organs of interest.

\subsection{Hyperparameter Analysis} \label{sec:s5_p3}
In this section, we conduct extensive analysis of several factors that correlate with segmentation performance of {\tt MOSformer}. Default configurations of {\tt MOSformer} are highlighted in \sethlcolor{light-gray}\hl{gray}.

\textbf{Momentum Coefficient.} The momentum coefficient, as described in Eq.~(\ref{eq:mom}), is an important hyperparameter in our model. We carry out detailed analysis on how $m$ affects the model performance, as shown in Fig.~\ref{fig:mom_coef}. Our empirical observations indicate a consistent decline in model performance with incremental increases in $m$. This suggests that maintaining feature consistency achieved through a relatively low momentum coefficient is advantageous. Specifically, a high momentum value (e.g., $m=0.9$) leads to a significant drop in segmentation performance, from $85.63\%$ to $83.49\%$ in DSC. In the extreme case of no momentum ($m=0$), the performance is nearly the worst. These findings reinforce our motivation for extracting distinguishable and consistent slice features.

\textbf{Neighborhood Slice Number.} Since the proposed {\tt MOSformer} is a 2.5D-based model, it requires neighborhood slices as additional inputs, as illustrated in Section~\ref{sec:s3}. Thus, the number of neighborhood slices ($s$) is an important hyperparameter. Table~\ref{table:slice_number} reports quantitative results for three different $s$ parameters. It can be observed that segmentation performance initially increases and then decreases with an increasing value of $s$. Evidently, information from inter-slice enables our model to perceive partial structures of 3D medical volumes. However, a peculiar phenomenon emerges: segmentation performance of the model with $s = 2$ is worse than that with $s = 1$.  Similar observations have been reported in~\cite{zhang2022bridging}.  One possible explanation is that the most valuable inter-slice information is derived from adjacent slices. Introducing non-adjacent slices may bring redundant information, which contributes negatively to model performance. Additionally, as $s$ increases, the computational costs of our model also escalate. Based on the above observations, $s = 1$ is the most practical choice for our model.

\begin{table}[htbp]
\caption{Effect of neighborhood slice number $s$ on the multi-organ segmentation (Synapse) and the automatic cardiac diagnosis challenge (ACDC) datasets. The best results are highlighted in \textbf{bold}.
}
\label{table:slice_number}
\centering
\renewcommand\arraystretch{1.2}
\renewcommand\arraystretch{1.2}
\resizebox{\linewidth}{!}{
\begin{tabular}{lll|ll}
\toprule
\multirow{2}{*}{Number} & \multicolumn{2}{c|}{Synapse} & \multicolumn{2}{c}{ACDC} \\ \cmidrule{2-5} & DSC (\%) $\uparrow$ & HD95 (mm) $\downarrow$ & DSC (\%) $\uparrow$ & HD95 (mm) $\downarrow$ \\ \midrule
$s=0$ & 83.73 {\color{mydarkgreen}(-1.90)} & 18.59 {\color{mydarkgreen}(+5.19)} & 91.71 {\color{mydarkgreen}(-0.48)} & 1.64 {\color{mydarkgreen}(+0.56)} \\
\rowcolor{gray!20} $s=1$ & \textbf{85.63} & \textbf{13.40} & \textbf{92.19} & \textbf{1.08} \\
$s=2$ & 84.95 {\color{mydarkgreen}(-0.68)} & 16.78 {\color{mydarkgreen}(+3.38)} & 91.91 {\color{mydarkgreen}(-0.28)} & 1.16 {\color{mydarkgreen}(+0.08)} \\ \bottomrule
\end{tabular}
}
\end{table}
\begin{table}[htbp]
\caption{Effect of multi-scale inter-slice fusion on the multi-organ segmentation (Synapse) and the automatic cardiac diagnosis challenge (ACDC) datasets. The best results are highlighted in \textbf{bold}.
}
\label{table:scale}
\centering
\renewcommand\arraystretch{1.2}
\renewcommand\arraystretch{1.2}
\resizebox{\linewidth}{!}{
\begin{tabular}{lll|ll}
\toprule
\multirow{2}{*}{Scale} & \multicolumn{2}{c|}{Synapse} & \multicolumn{2}{c}{ACDC} \\ \cmidrule{2-5} & DSC (\%) $\uparrow$ & HD95 (mm) $\downarrow$ & DSC (\%) $\uparrow$ & HD95 (mm) $\downarrow$ \\ \midrule
$/16$ & 83.00 {\color{mydarkgreen}(-2.63)} & 21.54 {\color{mydarkgreen}(+8.14)} & 91.63 {\color{mydarkgreen}(-0.56)} & 1.14 {\color{mydarkgreen}(+0.06)} \\
$/8,/16$ & 83.76 {\color{mydarkgreen}(-1.87)} & 20.73 {\color{mydarkgreen}(+7.33)} & 91.75 {\color{mydarkgreen}(-0.44)} & 1.08 {\color{mydarkgreen}(+0.00)} \\
$/4,/8,/16$ & 84.52 {\color{mydarkgreen}(-1.11)} & 15.81 {\color{mydarkgreen}(+2.41)} & 91.94 {\color{mydarkgreen}(-0.25)} & 1.08 {\color{mydarkgreen}(+0.00)} \\
\rowcolor{gray!20} $/2,/4,/8,/16$ & \textbf{85.63} & \textbf{13.40} & \textbf{92.19} & \textbf{1.08} \\
\bottomrule
\end{tabular}
}
\end{table}
\begin{table}[htbp]
\caption{Model parameters, floating-point operations per second (FLOPs), and the average time required for segmenting individual cases. The input size of 2(.5)D-based and 3D-based models are set to $224\times224$ and $96\times96\times96$, respectively. $^*$ means the experiments are conducted on the \textit{test} set of the multi-organ segmentation (Synapse) dataset and repeated five times.
}
\label{table:complexity}
\centering
\renewcommand\arraystretch{1.2}
\renewcommand\arraystretch{1.2}
\resizebox{\linewidth}{!}{
\begin{tabular}{llccc}
\toprule
Dimension & Method & \#params (M) & FLOPs (G) & Time$^{*}$ (s) \\ \midrule
\multirow{3}{*}{2D} & UNet \cite{ronneberger2015u} {\tiny \color{gray}[MICCAI'15]} & 17.26 & 30.74 & 0.67\\
 & TransUNet \cite{chen2024transunet} {\tiny \color{gray}[MedIA'24]} & 93.23 & 24.73 & 5.69\\ 
 & MISSformer \cite{huang2023missformer} {\tiny \color{gray}[TMI'23]} & 35.45 & 7.28 & 7.20 \\ \midrule
\multirow{2}{*}{3D} & UNETR \cite{hatamizadeh2022unetr} {\tiny \color{gray}[WACV'22]} & 92.62 & 82.63 & 5.39 \\
 & nnFormer \cite{zhou2023nnformer} {\tiny \color{gray}[TIP'23]} & 149.13 & 246.10 & 10.13 \\ \midrule
\multirow{2}{*}{2.5D} & CAT-Net \cite{hung2022cat} {\tiny \color{gray}[TMI'22]} & 220.16 & 121.83 & 21.34 \\
 & \textbf{MOSformer {\tiny \color{gray}[Ours]}} & 77.09 & 100.06 & 5.10 \\ \bottomrule
\end{tabular}
}
\end{table}

\textbf{Multi-scale Inter-slice Fusion.} Multi-scale learning enables deep models to capture global spatial information and local contextual details. This conclusion has been supported by many studies~\cite{chen2024transunet},~\cite{cao2022swin},~\cite{you2022class}. In this paper, we further investigate multi-scale learning by incorporating inter-slice fusion. Table~\ref{table:scale} presents results derived from four different inter-slice fusion configurations. Our default model achieves significant performance improvements, such as $+1.11\%\sim+2.63\%$ gains in DSC on the multiorgan segmentation (Synapse) dataset and $+0.25\%\sim+0.56\%$ gains in DSC on the automatic cardiac diagnosis (ACDC) dataset. With more scales of inter-slice information fused, {\tt MOSformer} demonstrates an enhanced ability to comprehend global shapes and anatomical details within segmentation targets. This enhancement facilitates precise localization of semantic regions, resulting in higher DSC, and accurate classification of category boundaries, reflected in smaller HD95.

\subsection{Model Complexity}
Table~\ref{table:complexity} presents a comparison of five medical image segmentation models with {\tt MOSformer} across various dimensions, including model parameters, floating-point operations per second (FLOPs), and the average time required for segmenting individual cases. {\tt MOSformer} maintains a relatively small size (77.09 M) compared with 3D-based and 2.5D-based models. Furthermore, {\tt MOSformer} exhibits an inference speed only half that of {\tt nnFormer}~\cite{zhou2023nnformer}, even surpassing 2D-based {\tt TransUNet}~\cite{chen2024transunet} and {\tt MISSformer}~\cite{huang2023missformer}. These results indicate {\tt MOSformer} can achieve a favorable trade-off between model complexity and segmentation performance.

\section{Conclusion} \label{sec:s6}
This study proposes a \underline{\tt MO}mentum encoder-based inter-\underline{\tt S}lice fusion trans\underline{\tt former} ({\tt MOSformer}) for stable and precise medical image segmentation. Dual encoders with a momentum update are able to guarantee both feature distinguishability and consistency, beneficial for inter-slice fusion. Besides, rich contexts can be captured via inter-slice self-attention in the IF-Trans module. The superior performance to state-of-the-art methods on three benchmarks has demonstrated {\tt MOSformer}'s effectiveness and competitiveness. It will be extended to other downstream medical analysis tasks in our subsequent works.

\printcredits

\section*{Acknowledgments}
This work was supported in part by the National Key Research and Development Program of China under Grant 2023YFC2415100, in part by the National Natural Science Foundation of China under Grant 62222316, Grant 62373351, Grant 82327801, Grant 62073325, Grant 62303463, in part by the Chinese Academy of Sciences Project for Young Scientists in Basic Research under Grant No. YSBR-104, in part by the Beijing Natural Science Foundation under Grant F252068, Grant 4254107, in part by Beijing Nova Program under Grant 20250484813, in part by China Postdoctoral Science Foundation under Grant 2024M763535, in part by the Postdoctoral Fellowship Program of CPSF under Grant GZC20251170 and in part by CAMS Innovation Fund for Medical Sciences (ClFMS) under Grant 2023-I2M-C\&T-B-017.

\section*{Declaration of competing interest}
The authors declare that they have no known competing financial interests or personal relationships that could have appeared to influence the work reported in this paper.

\section*{Data availability}
Data will be made available on reasonable request.

\bibliographystyle{model1-num-names}

\bibliography{cas-refs}

\end{document}